\documentclass[draft]{agujournal2019} 
\usepackage{url} 
\usepackage{lineno}
\usepackage[inline]{trackchanges} 
\usepackage{soul}
\journalname{JGR: Space Physics}

\begin{document}

	\title{Low-cost Monitoring of Energetic Particle Precipitation: Weather Balloon-borne Timepix Measurements During the May 2024 Superstorm}

	\authors{L.~Olifer\affil{1}, P.~Manavalan\affil{1}, D.~Headrick\affil{1}, S.~Palmers\affil{1}, B.~Harbarenko\affil{1}, J.~Cai\affil{1}, J.~Fourie\affil{1}, O.~Bauer\affil{1}, I.~Mann\affil{1}}
	
	\affiliation{1}{Department of Physics, University of Alberta, Edmonton, AB, Canada}

	\correspondingauthor{Leonid Olifer}{olifer@ualberta.ca}
	
	\begin{keypoints}
		\item This study presents a low-cost, low-mass solution for high-fidelity analysis of electron precipitation using a balloon-borne X-ray detector
		\item Balloon measurements during the May 2024 superstorm captured a periodic bremsstrahlung X-ray flux structure, linked to ULF wave modulation
		\item The Timepix-based payload and developed radiation identification algorithm proved effective in detailed atmospheric radiation measurements
	\end{keypoints}

	\begin{abstract}
		Understanding energetic electron precipitation is crucial for accurate space weather modeling and forecasting, impacting the Earth's upper atmosphere and human infrastructure. This study presents a low-cost, low-mass, and low-power solution for high-fidelity analysis of electron precipitation events by measuring the resulting bremsstrahlung X-ray emissions. Specifically, we report on results from the flight of a radiation detector payload based on a silicon pixel read-out Timepix detector technology, and its successful utilization onboard a `burster' weather balloon. We launched this payload during the May 2024 superstorm, capturing high-resolution measurements of both background galactic cosmic ray radiation as well as storm-time energetic electron precipitation. We further developed particle and radiation detection algorithms to separate bremsstrahlung X-rays from other particle species in the pixel-resolved trajectories as seen in the Timepix detector. The measurements revealed a distinctive four-peak structure in X-ray flux, corresponding to periodic four-minute-long bursts of energetic electron precipitation between 21:20 and 21:40 UT. This precipitation was also observed by a riometer station close to the balloon launch path, further validating balloon measurements and the developed X-ray identification algorithm. The clear periodic structure of the measured precipitation is likely caused by modulation of the electron losses from the radiation belt by harmonic Pc5 ULF waves, observed contemporaneously on the ground. The study underscores the potential of compact, low-cost payloads for advancing our understanding of space weather. Specifically, we envision a potential use of such Timepix-based detectors in space science, for example on sounding rockets or nano-, micro-, and small satellite platforms. 
	\end{abstract}

    \section{Introduction}
        Over 65 years ago, the discovery of the hazardous trapped electron population in the Van Allen radiation belts sparked significant efforts to understand the physical processes behind their formation and dynamics \cite<e.g.,>[]{VanAllen_1959}. More recently, the physical processes coupling the solar wind and the terrestrial magnetosphere with the thermosphere and the climate of our planet have come into focus \cite<e.g., see reviews by>[]{Heelis_2014, Sarris_2019}. Electron precipitation is one of the fundamental space weather processes in the coupled Magnetosphere-Ionosphere-Thermosphere (MIT) system and has been shown to have significant potential effects on both the natural environment and human infrastructure \cite<e.g.,>[and references therein]{Codrescu_1997, Rozanov_2012, Randall_2015, Mironova_2015, Marshall_2020}. For example, energetic electron precipitation from the Van Allen radiation belts can alter the chemical composition of the upper atmosphere by creating odd NOx and HOx species, which are known to be catalytic destroyers of ozone, thus, potentially impacting the Earth's weather and the climate \cite<e.g.,>[]{Seppala_2007, Nesse_2010, Sinnhuber_2012, Nesse_2016}. Energetic electron precipitation can also affect the ionosphere, leading to a cascade of ionization processes, resulting in potential disruption to communication and GPS systems \cite{Crowley_2018, Yang_2020}. 

        Therefore, a better understanding of processes causing energetic electron precipitation is crucial for improved space weather modeling and forecasting. More generally and as described above, energetic electron precipitation plays a key role in the energy budget of the Earth's upper atmosphere, affecting atmospheric chemistry and dynamics \cite<e.g.,>[]{Rozanov_2012}. Especially interesting are the processes that cause the most intense energetic electron precipitation during geomagnetic storms. In particular, the interaction of Van Allen radiation belt electrons with various plasma waves in the magnetosphere can cause different types of energetic particle precipitation into the atmosphere. For example, chorus and hiss waves can cause precipitation of electrons with a wide range of energies between a few 10s of keV to multiple MeV \cite<e.g., recent works by>[and references therein]{Chen_2021, Chakraborty_2022, Ma_2021}. Chorus waves are also theorized to be the driver for the very short (hundreds of ms) but very intense intervals of energetic electron precipitation called `microbursts' \cite<e.g.,>[]{Hikishima_2010}. Meanwhile, electromagnetic ion cyclotron (EMIC) waves can cause relativistic ($>\sim$1~MeV) electron precipitation \cite<e.g.,>[and references therein]{Usanova_2014, Grach_2022}. Ultra-low frequency (ULF) waves have also been shown to modulate energetic electron precipitation \cite<e.g.,>[]{Watt_2011, Shang_2021}. Importantly, earlier studies have shown that the overall energy input into the atmosphere from energetic electron precipitation during the few days of a geomagnetic storm is typically higher than the integrated input from more extended periods of lower intensity precipitation in between the storms \cite{Olifer_2023}. 

        Data from satellite missions has already significantly advanced our understanding of energetic electron precipitation. Low Earth Orbit (LEO) satellite missions such as the Polar Operational Environmental Satellites \cite<POES, e.g.,>[]{Peck_2015}, the Electron Losses and Fields Investigation \cite<ELFIN,>[]{Angelopoulos_2020}, and Focused Investigations of Relativistic Electron Burst Intensity, Range, and Dynamics II \cite<FIREBIRD II,>[]{Johnson_2020} missions have all provided valuable data on the spatial and temporal characteristics of energetic electron precipitation. In particular, data from the POES satellites has been instrumental in providing a long-term data set that has enabled statistical characterization of precipitation as a major contribution to space weather research \cite<e.g.,>[]{Rodger_2013, Yahnin_2016, Ozeke_2024}. The ELFIN mission, on orbit between 2018 and 2022, has enhanced our understanding of the physical processes contributing to energetic electron losses by capturing high-resolution measurements of electron pitch angle distributions during storm times \cite<e.g.,>[]{Zhang_2022, Mourenas_2024}. Meanwhile, FIREBIRD II data allowed for a more detailed investigation of the energy dependence and spatial scales of microburst particle losses \cite<e.g.,>[]{Shumko_2018, Kawamura_2021}. 

        Stratospheric balloon campaigns, such as the Balloon Array for Radiation-belt Relativistic Electron Losses \cite<BARREL,>[]{Millan_2013}, have complemented satellite observations by providing atmospheric measurements of precipitating particles. BARREL campaigns have been particularly effective in assessing high-energy electron precipitation events contemporary on multiple balloons at various spatial scales and on a wide range of timescales (from milliseconds to hours) associated with geomagnetic storms. They have also been used to examine connections between energetic electron precipitation and the inner-magnetosphere drivers \cite<e.g.,>[]{Woodger_2015, Cantwell_2024}. Note that unlike LEO satellites, the balloons perform indirect observations of energetic electron precipitation by measuring bremsstrahlung X-rays produced by energetic electrons braking in the atmosphere \cite<cf. e.g.,>[and references therein]{Foat_1998, Xu_2019}.  

        Typically, high-altitude balloon measurements of bremsstrahlung X-rays have been performed using detectors based on a scintillator crystal (e.g., NaI) and a photomultiplier tube (PMT). Such payloads allow for a detailed characterization of the X-ray energy spectra and can operate at a high temporal cadence. However, scintillator crystal-based spectrometers are typically heavy (a few kilograms) and therefore can only be used on higher-volume balloons. For example, BARREL flights have been undertaken using zero-pressure balloons carrying 20~kg payloads and remaining in the air for multiple days \cite{Millan_2013}. Such operations incur high costs and prevent frequent balloon campaigns.

        In this paper, we present a low-cost, low-mass, and low-power solution for performing measurements of bremsstrahlung X-rays, as well as other types of atmospheric radiation, using a silicon pixel read-out detector Timepix \cite{Llopart_2007} on a `burster' weather balloon. While such balloons stay in the stratosphere for only a few hours, they are far more accessible in terms of cost and the training required for their operations further enabling their utilization by a team of undergraduate students. We further describe the design of the payload and the data analysis techniques developed and implemented for interpreting the Timepix radiation data in the context of high-altitude weather balloon observations. Additionally, we present real flight measurements taken during the May 2024 superstorm and reveal periods of intense energetic electron precipitation in both balloon and riometer data combined with further evidence for ULF modulation of the precipitation from ground-based magnetometers. These observations additionally validate the developed X-ray detection algorithm in the Timepix-type chips. Overall, we demonstrate that modern Timepix pixelated particle detectors offer a cost-effective solution for frequent space weather measurements from high altitudes. By hosting these detectors on short-duration, weather `burster' balloon platforms, we can achieve similar results at a lower cost compared to conventional heavy-weight scintillator-based detectors, which require much larger balloons.
               

    \section{Instrumentation, Methodology, and Flight}
\begin{figure}
            \centering
            \makebox[\textwidth][c]{\includegraphics[width=1.4\linewidth]{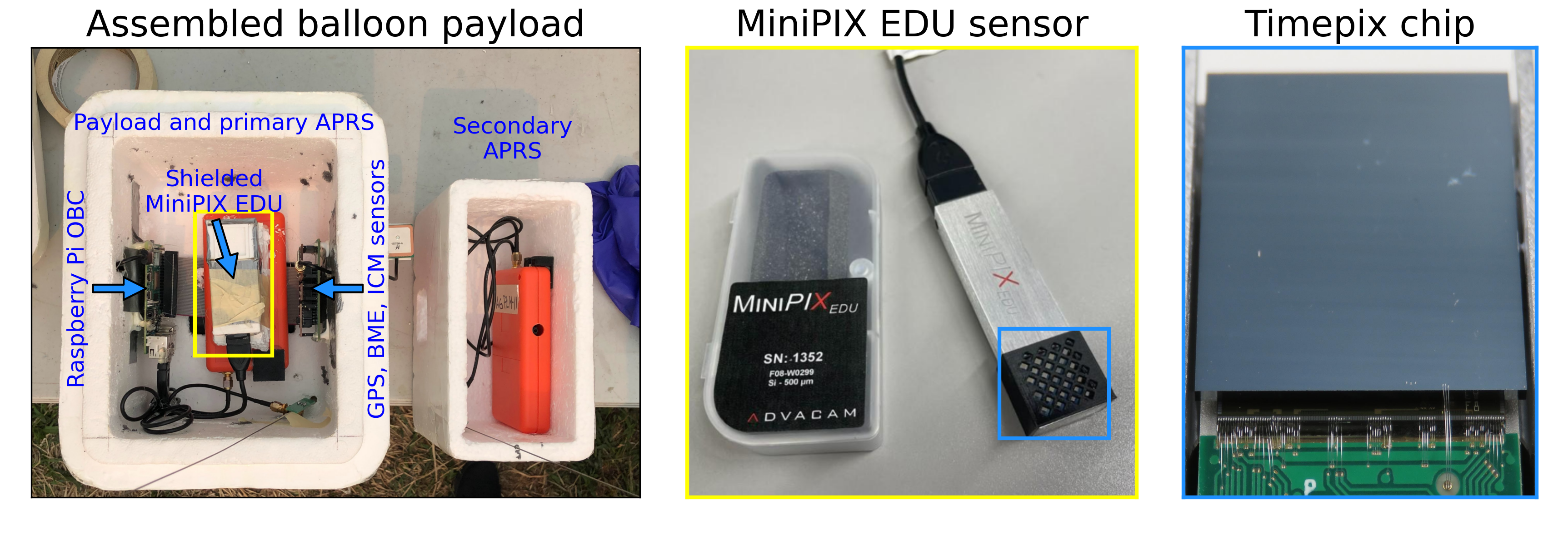}}
            \caption{The components of the balloon payload used for atmospheric radiation and X-ray bremsstrahlung measurements in this experiment. (Left panel) The assembled balloon payload includes Raspberry Pi on-board computer (OBC); the shielded Timepix-based MiniPIX EDU sensor; a suite of GPS, temperature and pressure (Adafruit BME280), and accelerometer (Adafruit ICM-20948) sensors, as well as the primary and secondary APRS (Automatic Packet Reporting System) units. (Center panel) The MiniPIX EDU sensor without shielding. (Right panel) Close-up of the 0.5~mm thickness silicone Timepix chip, which comprises the core of the MiniPIX EDU sensor.}
            \label{fig:Payload}
        \end{figure}

        The balloon payload was developed in the space physics group at the University of Alberta, further supported by a University of Alberta undergraduate student team (AlbertaSat) who have a long-standing heritage in space mission development, including two successfully launched CubeSats Ex-Alta~1 \cite{Mann_2020} and Ex-Alta~2. Figure~\ref{fig:Payload} illustrates the components of the balloon payload used for the atmospheric radiation measurements. The left panel shows the assembled payload, which includes a Raspberry Pi on-board computer (OBC) for data processing and control, primary and secondary Automatic Packet Reporting System (APRS) units for tracking and recovery, with a suite of sensors including GPS for timing and location tracking, an Adafruit BME280 sensor for temperature and pressure measurements, and an Adafruit ICM-20948 sensor for 9-axis accelerometer data. The primary scientific payload on the balloon is a Timepix-based radiation detector MiniPIX EDU (Figure~\ref{fig:Payload} center panel) developed by Advacam, and capable of measuring various types of ionizing radiation including photons, electrons, and ions. This MiniPIX EDU is equipped with a Timepix silicone pixel detector of the first generation (Figure~\ref{fig:Payload} right panel), based on 500~$\mu$m thick silicone layer separated into 256$\times$256 pixels with a $\sim$14mm$\times$14mm sensitive area (with 55$\mu$m$\times$55$\mu$m pixel size). 
        
        The Timepix chip functionality inside the MiniPix EDU is similar to that of a charge-coupled device (CCD) in conventional cameras and it captures data in each pixel in a manner analogous to bubble chambers. During a single 0.3-second exposure during the balloon flight, the chip records the resulting tracks produced by ionizing radiation particles and the corresponding energy deposition in its matrix of square pixels while operating in a Time-over-Threshold (ToT) mode. This mode measures the energy losses of particles, including photons, as they interact with the pixels in the sensor. By mapping these interactions, the Timepix chip provides detailed information on the particle's path and energy characteristics. The ToT mode is calibrated by the supplier (Advacam) to provide energy readings in each pixel based on the time interval in which the pulse from a particle remains above a threshold. The shape of the tracks and the energy deposited in each pixel enables the characterization of particle types like X-rays, electrons, protons, or ions. See Section~3., Data Analysis, for more information. 
        
        During the flight, the MiniPIX EDU was placed in a 3D-printed harnessing with 1~mm lead shielding in the walls to minimize sideways or backward particle penetration of the detector, effectively limiting its field of view to the 2$\pi$~sr above the silicone detector surface. In this case, it is also possible to estimate the geometric factor, $G$, of such an instrument following the derivation for a single plane detector of area, $A$, sensitive to particles coming from one side only \cite{Sullivan_1971}. The resulting $G=\pi A=$6.23~cm$^2$sr. This can be further used to estimate the flux, $J$, of charged particles and X-rays from recorded count rates of each separate species, $N$, in physical units of cm$^{-2}$s$^{-1}$sr$^{-1}$ as $J=N/G$. Table~1 presents additional technical details on the MiniPIX EDU chips and the balloon payload used in this experiment.  

        \begin{table}[]
        \caption{Technical characteristics of the MiniPIX EDU sensor and the balloon-borne payload used in the experiment}
        \label{tab:tab_S1}
        \centering
        \begin{tabular}{ll}
        \hline\hline
        \multicolumn{2}{c}{MiniPIX EDU sensor}          \\ \hline
        Readout chip type       & Timepix (gen 1)       \\
        Pixel size              & 55x55~$\mu$m             \\
        Sensor size             & 14x14~mm              \\
        Sensor resolution       & 256x256~pixels        \\
        Sensor thickness        & 500~$\mu$m               \\
        Interface               & USB                   \\
        Power draw              & 2.5~W                 \\
        Mass                    & 30~g                  \\
        Operating exposure time & 0.3~sec               \\ \hline
        \multicolumn{2}{c}{Balloon payload and mission} \\ \hline
        Balloon type            & 1200g weather balloon \\
        Total payload mass      & 780~g                 \\
        Power source            & Anker PowerCore 10000 mAh\\
        On board computer       & Raspbery Pi 4         \\
        On board storage        & 32 GB SD card         \\ 
        Maximum altitude reached& 32.6~km               \\ 
        Total flight time       & 4~hr 40~min           \\ \hline
        \end{tabular}
        \end{table}

\begin{figure}
            \centering
            \makebox[\textwidth][c]{\includegraphics[width=1\linewidth]{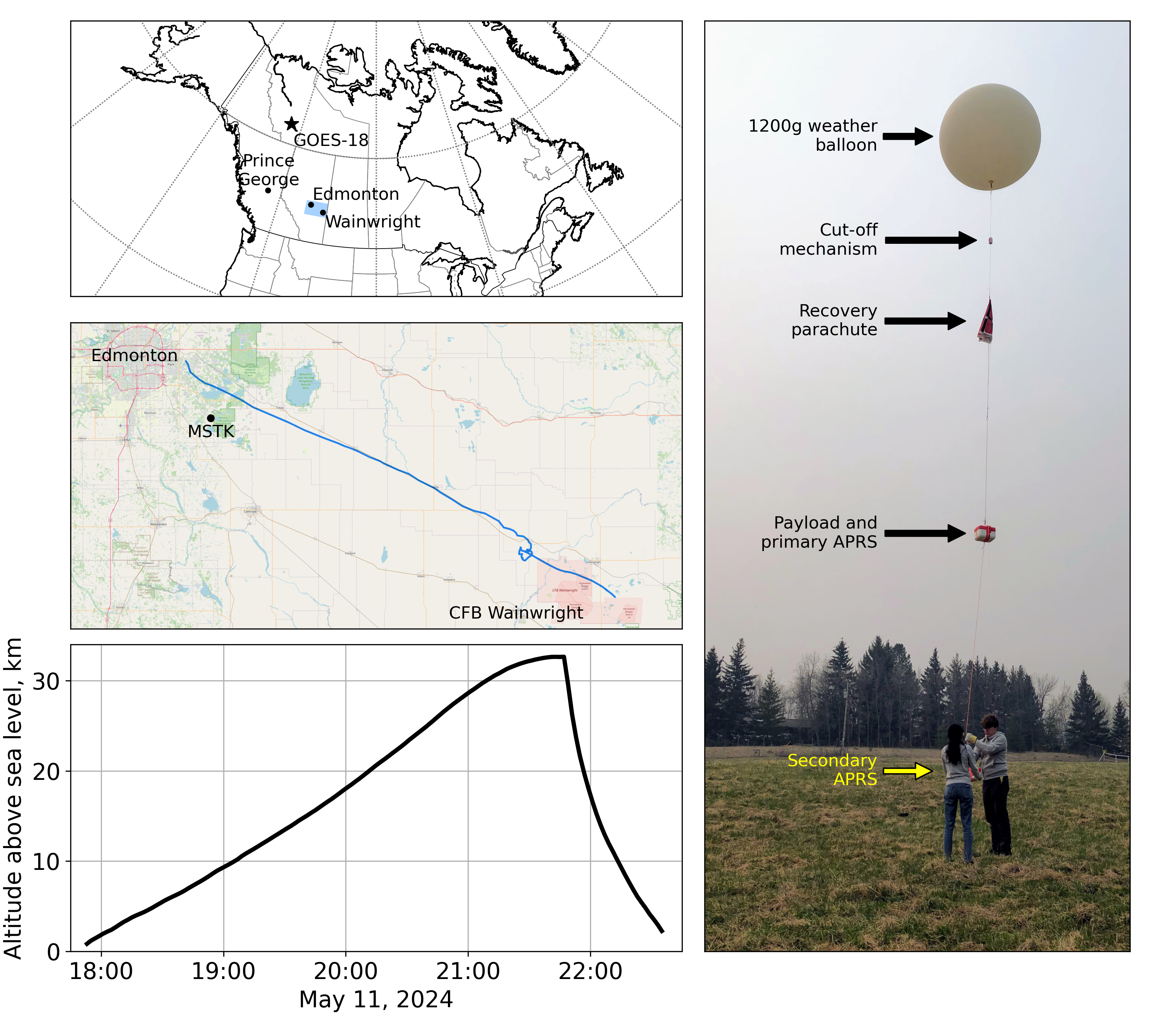}}
            \caption{Summary of the balloon flight campaign characteristics. (Top left and center left panels) The launch and flight path indicated on a map of Canada and a detailed map of part of Alberta, respectively. Both maps show key locations including the flight origin in Edmonton, the retrieval location at Canadian Forces Base (CFB) Wainwright, as well as the Ministik (MSTK) ground magnetometer and Prince George riometer stations. The top left panel map shows the area of the center panel map with a blue rectangle. (Bottom left panel) The altitude profile of the balloon flight shows the ascent to approximately 33~km above sea level, followed by descent. (Right panel) A photo of the balloon launch, illustrating the balloon setup, which includes a 1200g weather balloon, a cut-off mechanism, a recovery parachute, and the payload with the primary and secondary APRS tracking units. The secondary APRS unit is visible in the hands of the team members during the pre-launch preparations shown in the image.}
            \label{fig:Mission}
        \end{figure}

        The small weather balloon was launched from Edmonton on May 11, 2024. It ascended to 33~km altitude over a period of $\sim$4~hr, followed by an $\sim$1~hr descent. The balloon payload landed on the territory of the 3rd Canadian Division Support Base Detachment Wainwright, also referred to as Canadian Forces Base Wainwright or CFB Wainwright. The payload was successfully found and retrieved from the training firing range by the team members with the support of Canadian military personnel. Figure~\ref{fig:Mission} illustrates the primary characteristics of the May 11, 2024, balloon flight campaign, detailing the launch and flight path, altitude profile, and balloon setup. The right panel of Figure~\ref{fig:Mission} features a photograph of the balloon being prepared for launch from Edmonton by L.~Olifer and P.~Manavalan.  

        During the 4-hour-and-45-minute flight, the primary MiniPix EDU scientific payload produced approximately 50,000 exposure files, at a cadence of 0.3~s, totaling 6.5 GB of raw data. The secondary temperature/pressure and accelerometer sensors operated as intended; however, the GPS unit failed in the early stages of the flight. Consequently, the payload measurements were timestamped by the internal Raspberry Pi clock, which was synchronized with GPS time information before the flight. The cut-off mechanism, designed to prevent the balloon from drifting too far, was programmed to sever the weather balloon from the parachute and payload at the three-hour mark. However, it also failed, causing the balloon to remain at high altitudes for approximately an additional hour before the balloon itself burst. As will be evident from the data presented in later sections of this paper, this extra hour of measurenments was crucial for identifying an energetic electron precipitation event in the bremsstrahlung X-ray flux data from the MiniPix EDU. This incident highlights that unexpected outcomes or partial payload failures can sometimes lead to valuable scientific observations.

        \begin{figure}
            \centering
            \makebox[\textwidth][c]{\includegraphics[width=0.8\linewidth]{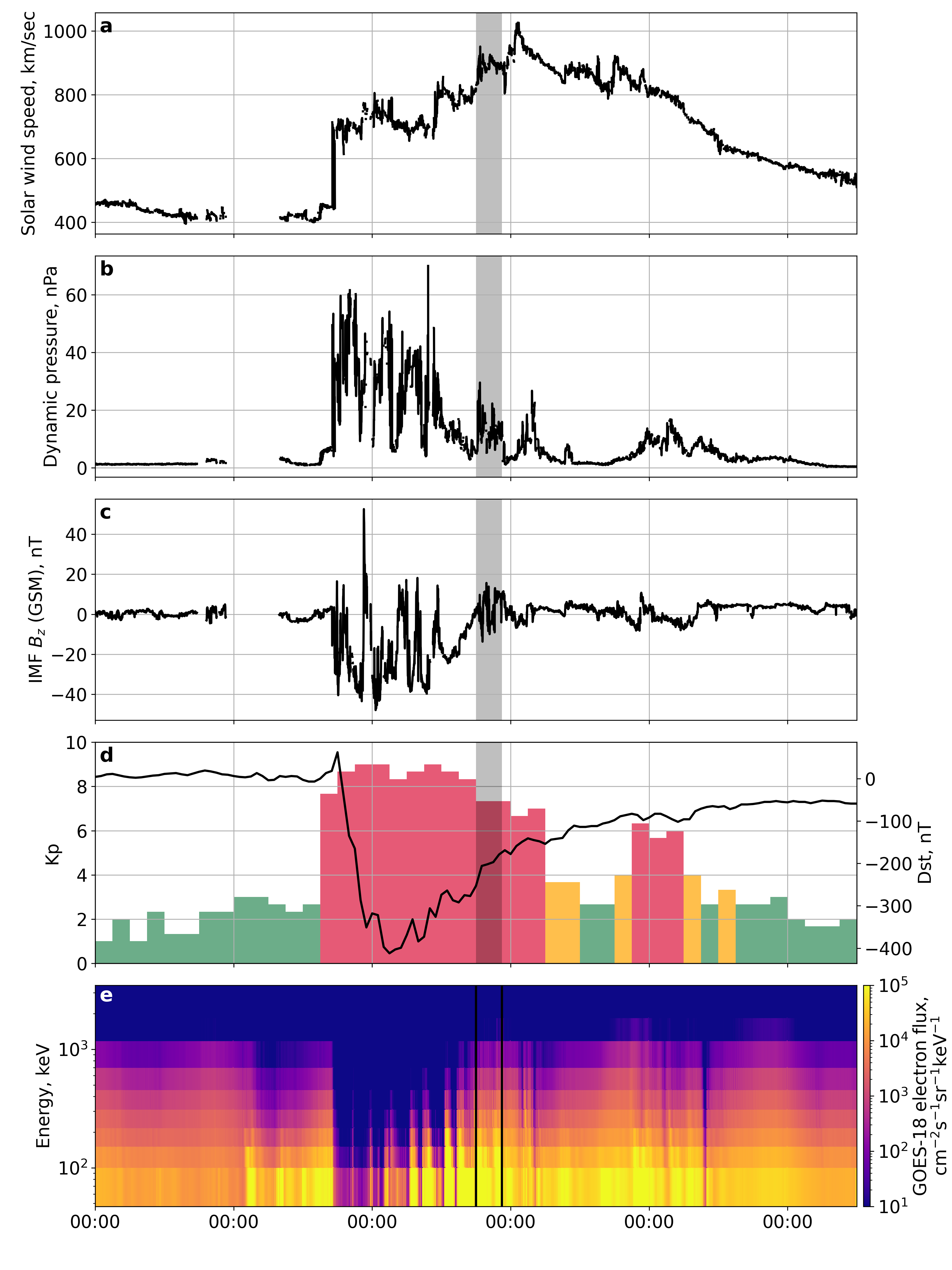}}
            \caption{Selected space weather conditions during the May 2024 geomagnetic storm. The panels from top to bottom display the following parameters: (a) Absolute value of solar wind speed, indicating a sharp increase corresponding to the start of the geomagnetic storm late on May 10, 2024. (b) Dynamic pressure of the solar wind. (c) The $B_z$ component of the interplanetary magnetic field (IMF) in GSM coordinates, highlighting strong negative values during the storm. (d) The Kp index (colored bars) and Dst index (black line) representing geomagnetic activity, with Dst reaching a minimum value of $-$412~nT. (e) GOES-18 differential electron flux at geosynchronous orbit across a range of energies, illustrating increased flux levels during the period following the minimum Dst. The shaded region indicates the duration of the balloon flight.}
            \label{fig:Storm}
        \end{figure} 
        
        The balloon launch was scheduled for 18:00 UTC on May 11, 2024, following the update of the NOAA 3-Day Space Weather Forecast released on May 10, which included a potential G5 storm. The team prepared the payload for launch and notified air traffic control on the morning of May 10 within 24 hours of the desired flight window, ensuring the launch would occur within the geomagnetically active period. The primary aim of the mission was to measure relativistic electron precipitation caused by EMIC waves in the noon/post-noon local time region during the storm recovery phase. Figure~\ref{fig:Storm} presents an overview of the solar wind and geomagnetic conditions during the May 2024 geomagnetic storm. This was the first G5 geomagnetic storm since the Halloween storms of 2004, characterized by a Kp index of 9 and a Dst index which reached a minimum of $-$412~nT. Notably, the IMF $B_z$ component, while reaching very low values around $-$42~nT, exhibited significant fluctuations, turning southward and northward on an hourly timescale. Electron flux measurements from GOES-18 reveal an initial loss of particles around 16:00~UT on May 10, followed by the recovery and acceleration of the electron population starting close to minimum Dst around 00~UT on May 11. The balloon was launched at 17:45 UT on May 11, during the middle of the storm's recovery phase. The flight period itself is indicated by the shaded regions in all panels of Figure~\ref{fig:Storm}, where the value of Dst had partially recovered to values of $\sim-$200~nT.

        Overall, the launch conditions for the May 11, 2024, balloon flight presented an ideal opportunity not only for measuring bremsstrahlung signals from precipitating energetic electrons but also for testing the utility of the Timepix-based payload for space weather research on a `burster' weather balloon. The next section describes the Timepix data generated by the payload and the algorithm designed and implemented for the analysis of its data. This is followed by the results section, including experimental verification of the MiniPix EDU performance and the successful detection of ULF wave modulated bremsstrahlung X-rays created by energetic electron precipitation.  
    

    \section{Data Analysis}
\begin{figure}
            \centering
            \makebox[\textwidth][c]{\includegraphics[width=1.4\linewidth]{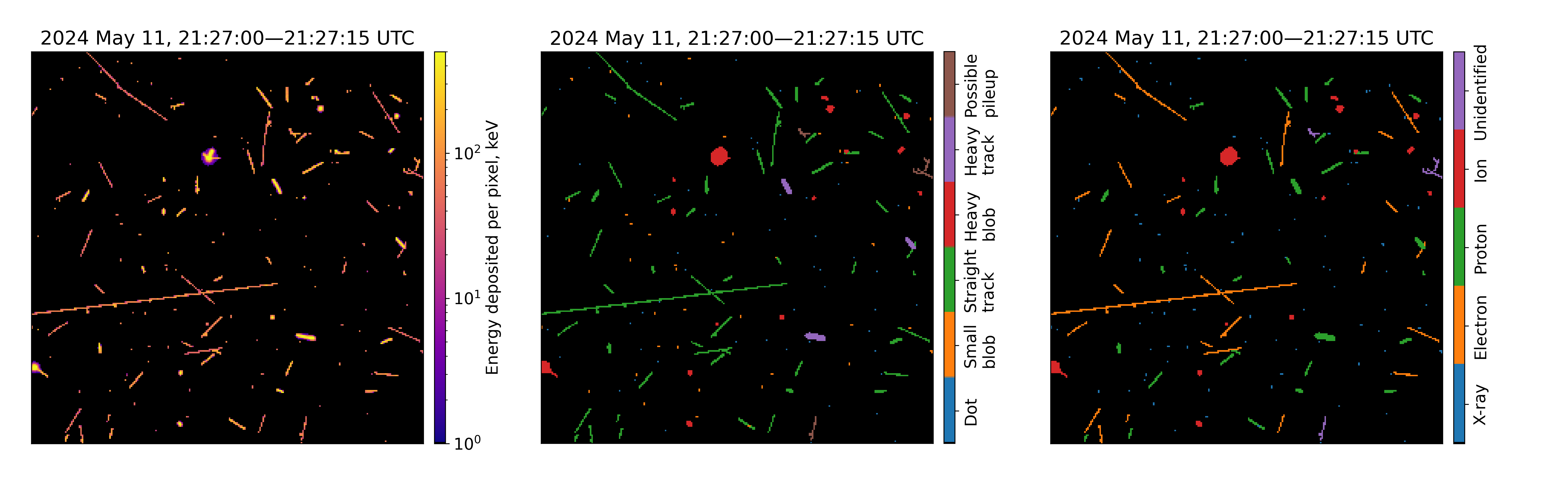}}
            \caption{An example of balloon flight MiniPIX EDU data processing during the balloon flight. This figure shows an example of the combined data from 45 frames for the period from 21:27:00 to 21:27:15~UT on May 11, 2024. For this study, we analyze each of the 50,000 frames separately and identify the species creating the observed tracks using a series of processing and selection algorithms. Each panel shows the combined characteristics from 15 seconds of 0.3~s individual exposure frames. (Left panel) Energy deposition per pixel in keV, indicating the particle and photon tracks. (Center panel) Particle track shape classification, distinguishing between different types of tracks such as dots, small blobs, straight tracks, heavy blobs, and potential pileups. (Right panel) Particle type identification, categorizing the detected particles into ions, protons, electrons, and X-rays based on the shape of the track and the total energy deposited as well as $dE/dx$ characteristics. See text for more details.}
            \label{fig:Timepix}
        \end{figure}

        The Timepix-based MiniPix EDU sensor demonstrated robust performance during the May 11, 2024, balloon flight, capturing high-resolution data on both energetic electron precipitation and galactic cosmic ray (GCR) background. Figure~\ref{fig:Timepix} presents an example of the data processing, showing the combined data from 45 frames of 0.3~s exposures for illustrative purposes. Note that it takes MiniPix EDU a variable amount of time (typically approximately 0.05~s) to save each of the exposure frames on the Raspberry Pi SD card, resulting in the typical raw data cadence of $\sim$0.35~s. For this study, we analyze each of the 50,000 0.3-second-long exposures separately, but superpose each of the analyzed frames from the 15~s interval in Figure~\ref{fig:Timepix}. The left panel of the figure illustrates the energy deposition per pixel in keV, forming clear tracks of ionizing radiation in the pixelated chip when superposed. These tracks are then grouped and classified based on their shape, following the naming convention from \citeA{Gohl_2019}. Each track is identified using a depth-first search (DFS) algorithm to find groups of neighboring pixels with non-zero energy deposition. While this method performs well for the chosen exposure duration during the balloon flight, it is incapable of identifying and distinguishing between potential pileups or low-energy electrons scattered inside the sensor. A more robust approach for these features could be utilized in the future. The center panel of Figure~\ref{fig:Timepix} shows the grouped and classified particle tracks, demonstrating the effectiveness of the DFS algorithm in track identification.
    
        In this study, we adopt a similar approach to that introduced by \citeA{Gohl_2019} for the classification of particle tracks into different radiation species in the Space Application of Timepix Radiation Monitor (SATRAM) payload data onboard the Proba-V satellite. For more specific details, refer to their section 2.2.3 and their Figure 8. We assign `dots' and `small blobs' to X-rays; `straight tracks' to electrons if their $-dE/dx \le 10$~MeVcm$^2$/g and to protons if their $-dE/dx > 10$~MeVcm$^2$/g; and `heavy blobs' and 'heavy tracks' to protons if their $-dE/dx \le 100$~MeVcm$^2$/g and to ions if their $-dE/dx > 100$~MeVcm$^2$/g. This approach slightly differs from \citeA{Gohl_2019} where X-rays were identified only for `dot'-type tracks. However, testing of our MiniPix EDU sensor in the laboratory with a known gamma source $^{133}$Ba shows that a single photon can deposit energy into multiple neighboring pixels (see Supplementary~Figure~S1). There is also a possibility that muons may be misidentified as electrons, and the particle classification algorithm could be additionally refined to incorporate additional GCR particle species. The right panel of Figure~\ref{fig:Timepix} shows the identified particle types resulting from the application of this identification scheme to the track types shown in the center panel. 

\begin{figure}
            \centering
            \makebox[\textwidth][c]{\includegraphics[width=0.5\linewidth]{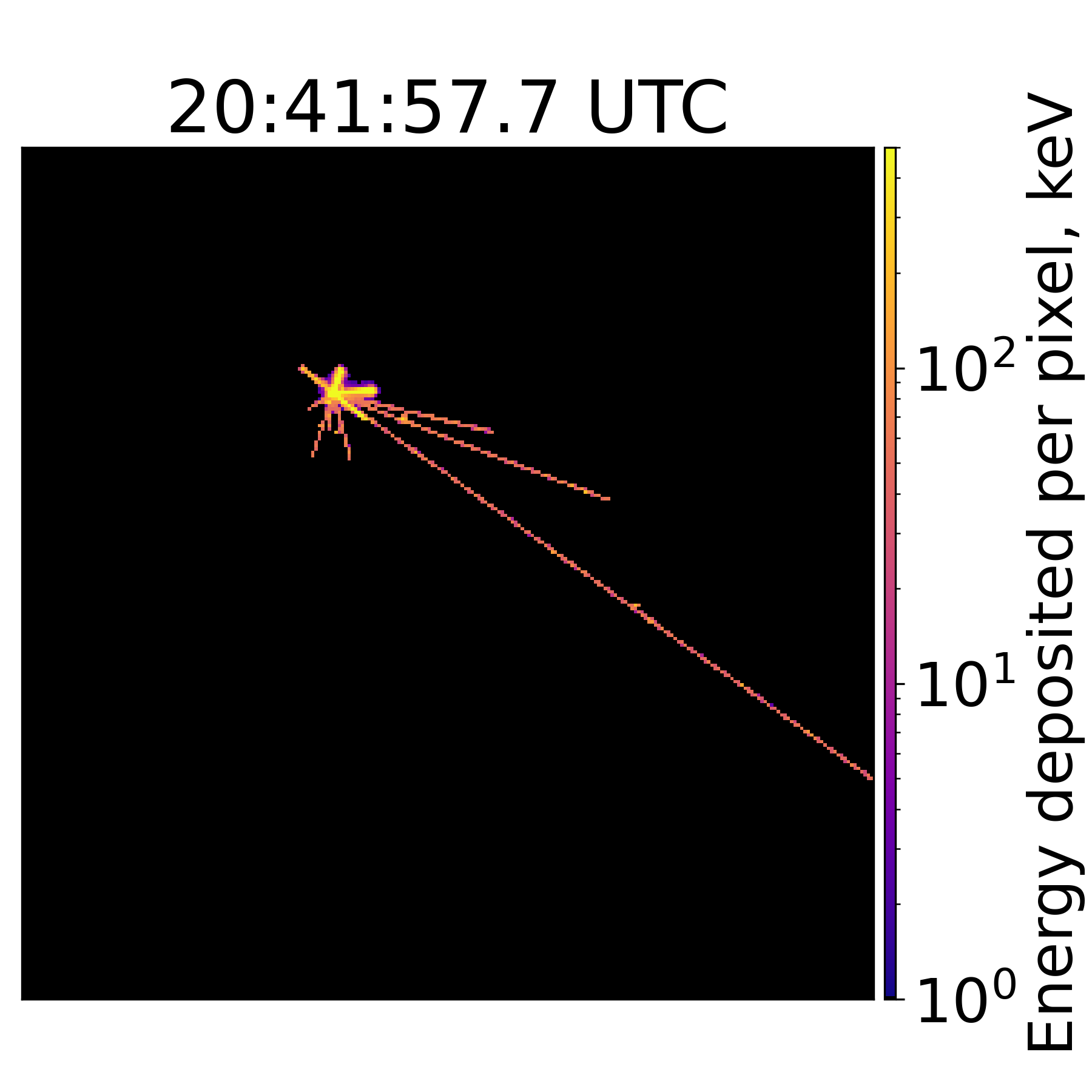}}
            \caption{An example of an unidentifiable track captured by the MiniPIX EDU sensor during the high altitude balloon flight at 20:41:57.7~UT on May 11, 2024. The complex shape and clearly discernible tracks of different thicknesses suggest a possible atomic or nuclear interaction producing multiple secondary particles of possible different species.}
            \label{fig:Nuclear}
        \end{figure}

        It is also important to note that the MiniPIX EDU data contained a number of unidentifiable tracks, which might be attributable to low-energy electrons scattered in the silicon detector or to pileup events (brown tracks in the center panel of Figure~\ref{fig:Timepix}), or to significant energy deposition events indicating possible atomic or nuclear interactions with multiple products. For example, Figure~\ref{fig:Nuclear} shows such an event for the single MiniPix EDU exposure frame recorded during the balloon flight. While such tracks are not the focus of this paper, they are intriguing nonetheless. These observations highlight the capabilities of the Timepix chips, which were originally developed for detailed particle tracking at CERN \cite{Llopart_2007}. The capability of these sensors to detect and record complex interaction events underscores their versatility and potential for a wide range of applications even in space weather research.

    \section{Results}
\begin{figure}
            \centering
            \makebox[\textwidth][c]{\includegraphics[width=1\linewidth]{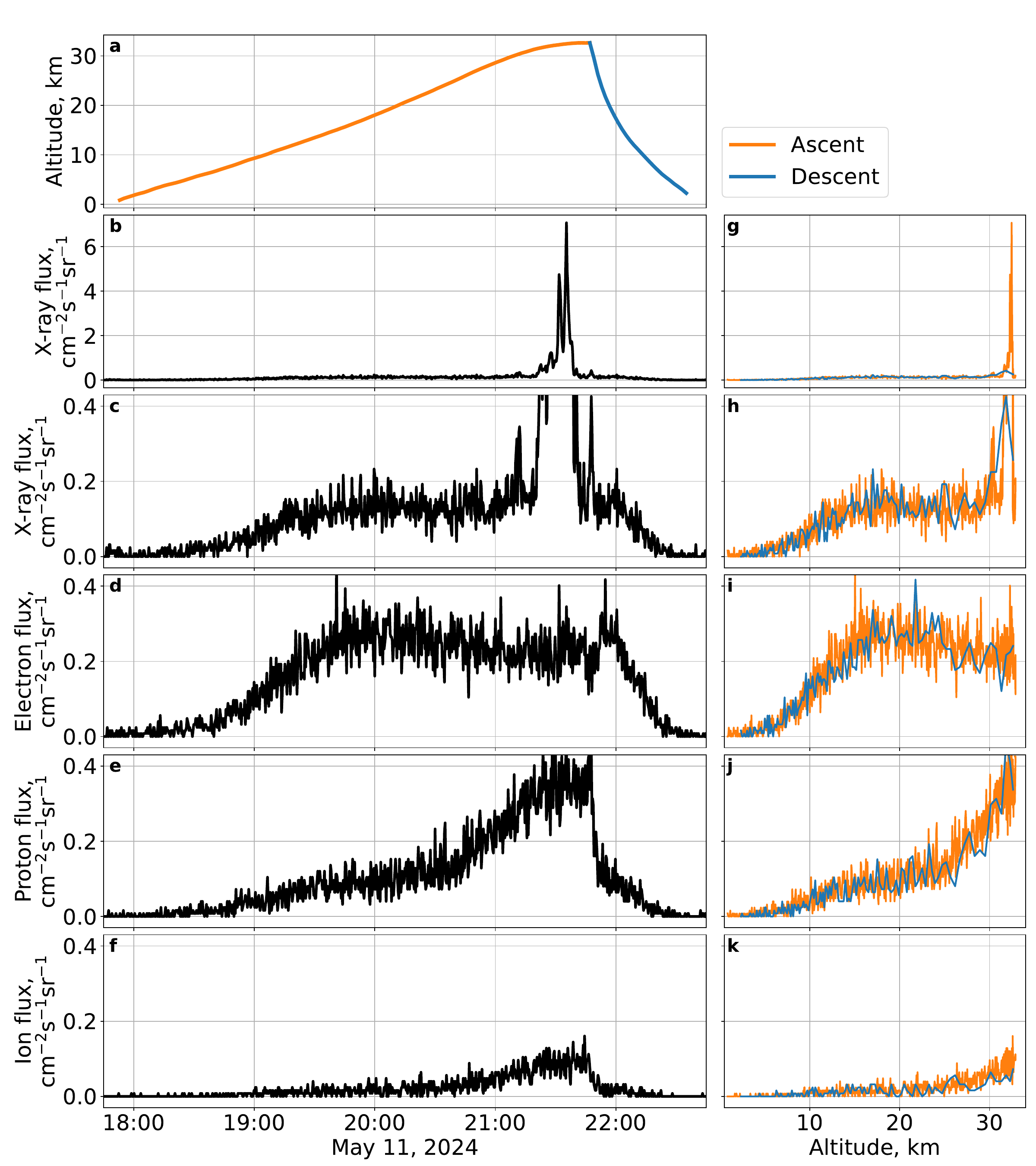}}
            \caption{Fluxes of classified particles observed by MiniPix EDU as a function of time and altitude during the May 11, 2024, balloon flight. The panels in the rows of each of the two columns show the fluxes of X-rays, electrons, protons, and ions, plotted as a function of UTC time (left column) and altitude (right column). The data captures both the ascent (shown in orange) and descent (shown in blue) phases of the balloon flight. Note that the X-ray flux is plotted in two rows to reveal the signal from both precipitating energetic electrons (larger flux in panels b and g) and the galactic cosmic ray (GCR) background (lower flux in panels c and h).}
            \label{fig:Fluxes}
        \end{figure}

        In this section, we summarize the results obtained during the balloon flight of the MiniPix EDU sensor and arising from the application of the species identification algorithm described in the previous section. Figure~\ref{fig:Fluxes}~(panels b-f) shows the measured integral fluxes of various classified particles and species as a function of time during the balloon flight. Figure~\ref{fig:Fluxes}(a) shows the altitude profile of the balloon, and Figure~\ref{fig:Fluxes}~(panels g-k) shows integral fluxes as a function of altitude, separated into ascent and descent stages by color. Each flux measurement shown in the figure incorporates information from approximately 30 frames, resulting in a fixed flux cadence of 10.0~s. Figure~\ref{fig:Fluxes}(b) shows the identified X-ray flux throughout the entire flight. It reveals a significant enhancement of X-ray flux with four distinct periodic peaks between 21 and 22 UT. The peaks have a distinct periodicity of $\sim$4~min, which is discussed later in the paper. Since the location of the balloon in latitude, longitude, and altitude does not change significantly during these observations (cf., Figure~\ref{fig:Mission}), we attribute these four peaks to temporal oscillations in the X-ray flux rather than to spatial changes caused by the balloon's movement.   
        
        Figure~\ref{fig:Fluxes} also indicates that the magnitude of the X-ray flux during these enhancements is significantly higher than the magnitude of the background measurements of the flux of various GCR particles (panels c through f), by a factor of 4 to 30. These periodic enhancements are also only visible in the X-ray flux, which we attribute to a different physical process than associated with the GCRs and specific to periods of storm-time periodic electron precipitation resulting in bremsstrahlung X-ray emissions (see Section~5 Discussion for more information). In this paper, we refer to the elevated fluxes in the X-ray data between 21:00 and 22:00 UT as the ``storm-time X-ray peak structure'' to highlight the distinct difference between these observations and the much smoother and lower flux characteristic of the background GCR radiation.

        Specifically, Figure~\ref{fig:Fluxes}~(panels c-f) and (h-k) show the measured X-ray, electron, proton, and ion fluxes plotted on the same y-axis range to reveal the GCR background fluxes. Both X-rays and electrons exhibit a characteristic increase in radiation flux with altitude until altitude of approximately 18-20~km, beyond which there is a slight decrease in these particle counts. This trend is observed during both the ascent and descent of the balloon (c.f., Figure~\ref{fig:Fluxes}~panels h and i). This peak in electron and X-ray flux is often referred to as the Regener-Pfotzer maximum \cite<c.f.,>[]{Carlson_2014}. The absolute values of the GCR background X-ray fluxes measured here are similar to those reported by \citeA{Ruffenach_2024} (their Figure 15). 
        
        Notably, previous studies have reported that the background X-ray flux produced by GCR is approximately an order of magnitude higher than that of electrons \cite{Lei_2004, Ruffenach_2024}. Meanwhile, in this experiment, the fluxes of GCR-related electrons and X-rays are comparable, with electron fluxes generally slightly higher than X-ray fluxes. Additionally, contrary to previous results, the proton flux in our experiment does not remain constant between approximately 20~and 30~km, as previously reported by \citeA{Lei_2004}. Instead, Figure~\ref{fig:Fluxes} showcases a monotonic increase in proton fluxes with altitude up to 33~km, instead of tapering off at around 20~km \cite{Lei_2004}. It remains unclear whether these discrepancies are due to the storm-time conditions of the event or the misidentification of non-accounted-for types of GCR particles (e.g., muons or pions) as electrons or protons. Nonetheless, the X-ray flux measured by the balloon mission reported here is generally representative and in good agreement with these prior studies. It confirms in general the validity of our photon observations and in particular the validity of our hypothesis that the periods of modulated large X-ray flux can be associated with the geomagnetic storm conditions, and specifically more likely with the precipitation of energetic electrons from the Van Allen radiation belts as a causative effect. 

\begin{figure}
            \centering
            \makebox[\textwidth][c]{\includegraphics[width=1\linewidth]{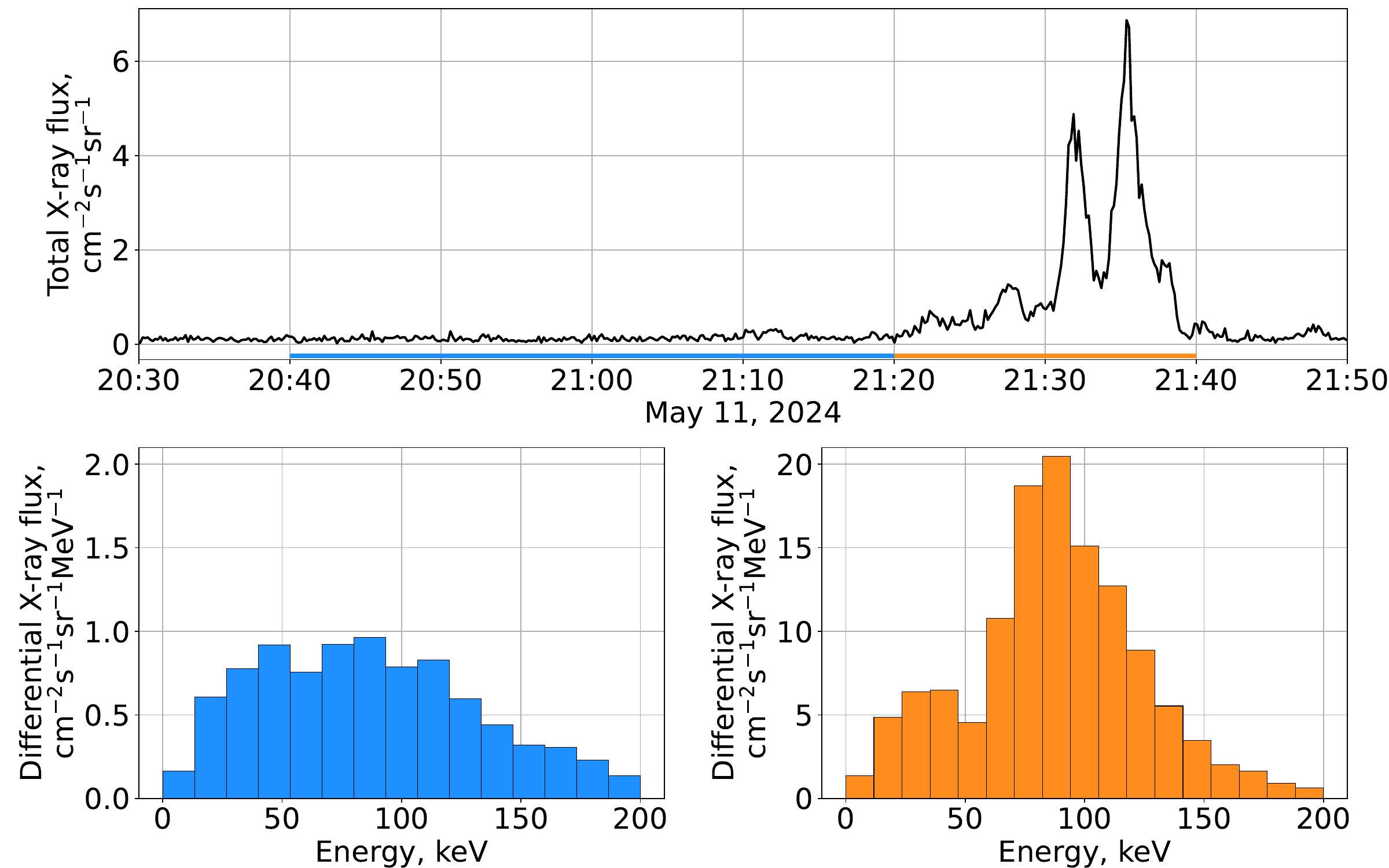}}
            \caption{Total and differential X-ray flux measurements during the May 11, 2024, balloon flight. (Top row) Total X-ray flux as a function of time, illustrating the temporal variations observed during the flight. The blue and orange bars at the bottom of the panel highlight the time periods for which the two energy spectra shown in the bottom row are calculated. (Bottom row) Differential X-ray flux as a function of energy (in keV) for the measurement periods observing the GCR background only (blue bar from the top panel; bottom left) and during the periods of the storm-time X-ray peak structure (orange bar from the top panel; bottom right).}
            \label{fig:E}
        \end{figure}

        It is also therefore interesting to investigate how the energy distribution of X-ray fluxes changes between the GCR background and those in the storm-time X-ray peak structure. Figure~\ref{fig:E} presents a detailed analysis of the X-ray flux during the final two hours of the balloon's ascent. Given the relatively low flux values, obtaining energy distributions for each frame is challenging. Consequently, we present only two energy spectra, each estimated over a 40-minute and a 20-minute period, respectively. These spectra are obtained firstly for the period observing very low GCR background fluxes $<$0.4~cm$^{-2}$s$^{-1}$sr$^{-1}$ (blue-shaded 40~min segment in Figure~\ref{fig:E}) and secondly in the storm-time X-ray peak structure (orange-shaded 20~min segment in Figure~\ref{fig:E}). The two energy spectra not only differ in their intensity but also in their energy spectral shape. The orange distribution (storm-time X-ray flux; bottom right panel of Figure~\ref{fig:E}) shows the existence of a second distinct higher energy X-ray population, clearly visible as a second peak in the differential flux as compared to the GCR interval (blue; bottom left panel of Figure~\ref{fig:E}).  

        The measurement of two characteristically different energy spectra in Figure~\ref{fig:E} further suggests that the processes responsible for the storm-time X-ray peak structure between 21:20 and 21:40 UT are distinct from those creating the GCR background. The enhanced X-ray flux at higher energies observed during the period of very high X-ray flux exhibits a markedly different energy distribution compared to the GCR differential flux, indicating that additional physical mechanisms are at play. Bremsstrahlung X-ray emissions from precipitating energetic electrons are a likely candidate for the creation of this enhanced and higher energy X-ray flux structure. Interestingly, such high X-ray flux has to our knowledge not been recorded previously by GCR-focused missions during quiet geomagnetic conditions, further highlighting storm-time electron precipitation as the likely cause of the observed intense four-peaked X-ray structure. In the next section, we further assess the hypothesis that energetic electron precipitation was the cause of the modulated and elevated X-ray flux observed by the balloon between 21:20 and 21:40~UT on May 11, 2024.
        

    \section{Discussion}
        \begin{figure}
            \centering
            \makebox[\textwidth][c]{\includegraphics[width=1\linewidth]{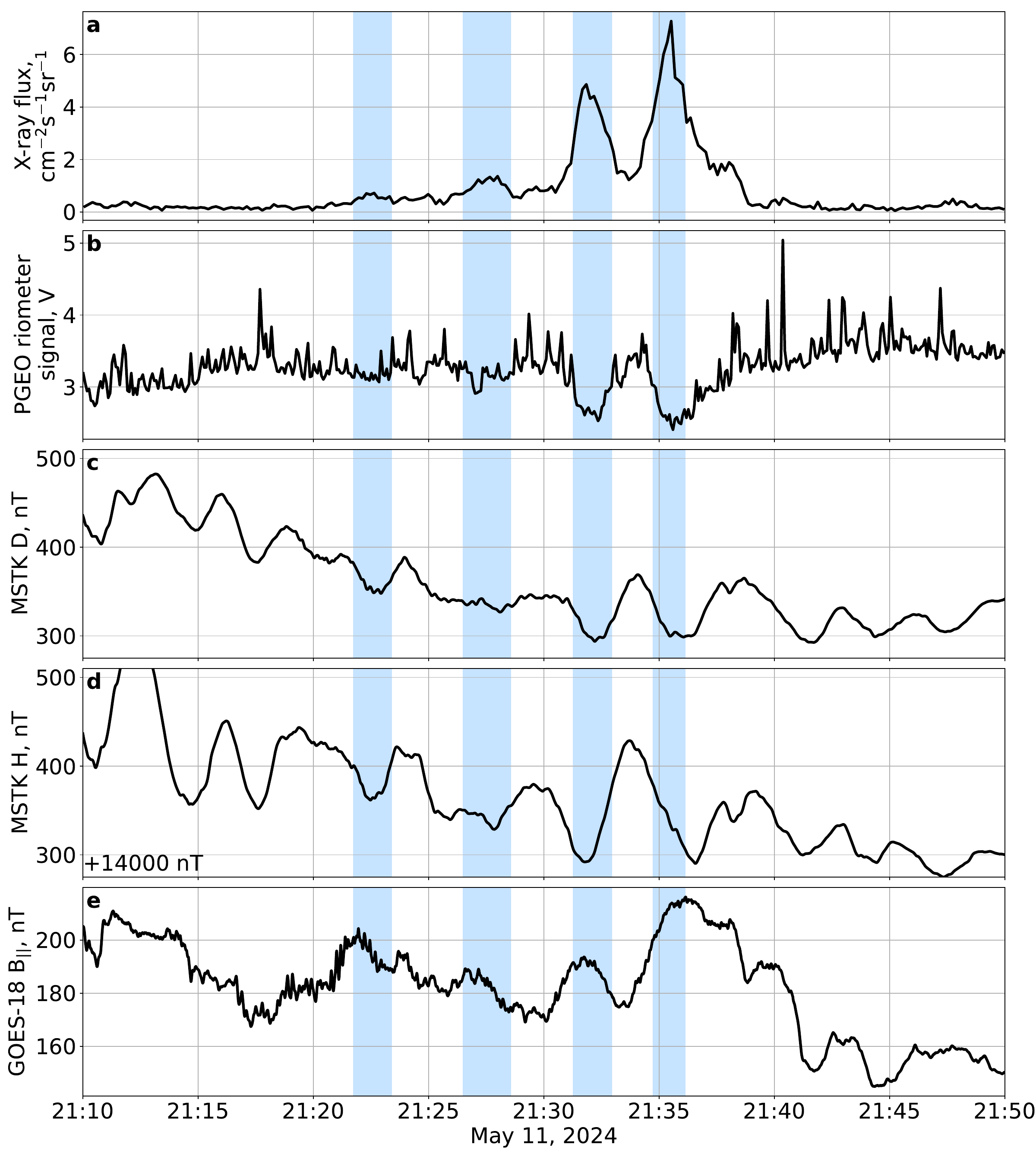}}
            \caption{Comparison of the storm-time X-ray peak structure with ground-based and satellite data. (a) X-ray flux exhibiting the storm-time X-ray peak structure observed at the balloon between 21:20 and 21:40 UT. (b) Prince George (PGEO) riometer signal, indicating modulated ionospheric radio absorption as an indirect measure of electron precipitation. (c and d) D and H components (magnetic East-West and North-South respectively) of the magnetic field measured by the Ministik Lake (MSTK) ground magnetometer station from CARISMA array. (e) GOES-18 parallel component of the magnetic field in the field-aligned coordinate system. The blue highlighted regions in each panel show the timing of the balloon X-ray peaks spanning their full width at half maximum.}
            \label{fig:precip}
        \end{figure} 
        
        Bremsstrahlung X-ray emissions from precipitating energetic electrons provides a natural explanation for the observed storm-time X-ray peak structure seen in the balloon observations. The four-peak structure suggests that the MiniPix EDU detected four distinct precipitation bursts, occurring with an approximately 4-minute period and with a progressively increasing flux. While energetic electron precipitation signatures are commonly observed at balloon altitudes in X-ray measurements, the four-peak structure is unusual. Therefore, it is valuable to compare these balloon observations with other ground-based and in-situ satellite data during the period of the observed storm-time X-ray peak structure to examine further the nature of the process that could be involved in creating the characteristic modulated energetic electron precipitation. Figure~\ref{fig:precip} provides a comparison of the storm-time X-ray peak structure with corresponding measurements from the Prince George (PGEO) riometer \cite<e.g.,>[]{Spanswick_2013}, the Ministik Lake (MSTK) ground magnetometer from Canadian Array for Realtime Investigations of Magnetic Activity \cite<CARISMA>[]{Mann_2008}, and the magnetometer on-board the GOES-18 satellite \cite{Singer_1996}. See Figure~\ref{fig:Mission} for locations, including the magnetic footprint of GOES-18. The X-ray flux in the top panel of Figure~\ref{fig:precip} shows the detailed time-series of the storm-time X-ray peak structure observed on the balloon between 21:20 and 21:40 UT. The blue highlighted regions in Figure 8 indicate the timing of these X-ray peaks, spanning their full width at half maximum.

        Importantly, the ground-based riometer signal for the PGEO station shown in panel Figure~\ref{fig:precip}b shows two distinct drops in voltage, indicative of increased ionospheric cosmic radio absorption, occurring at the time of the last two peaks in the X-ray measurements. This riometer observation is characteristic of energetic electron precipitation with energies $>\sim$30~keV \cite<e.g.,>[]{Rodger_2013}. The contemporaneity of the enhanced X-ray flux and the proximally observed riometer absorption, as well as the temporally increasing signal intensity in both riometer and balloon data, strongly suggests that the balloon payload observed bremsstrahlung X-ray emissions arising from precipitating energetic electrons. Interestingly, the balloon data has lower noise than the ground-based riometers, allowing it to be able to resolve the much weaker precipitation signals during the first two storm-time X-ray peaks, but which are more difficult to see in the PGEO riometer data.

        Figure~\ref{fig:precip}c-e also show magnetic field measurements from the ground-based magnetometer at Ministik Lake, near Edmonton from the CARISMA array, and from GOES-18 in geosynchronous orbit (see Figure~\ref{fig:Mission}). A clear Pc5 harmonic oscillation in the magnetic field is observed both on the ground and at GOES-18 in space, with the same $\sim$4-minute periodicity as the observed precipitation peaks in the balloon and riometer data, although the phase relationship appears to be more complex. The role of ULF waves in either directly driving electron precipitation \cite{Rae_2018}, or modulating it \cite<e.g.,>[]{Spanswick_2005, Watt_2011, Brito_2015, Shang_2021}, have been studied in the past. These observations suggest that the Pc5 waves could have played a potentially significant role in structuring the periodic precipitation events observed during the balloon flight.  

        A more detailed analysis of this intriguing periodic precipitation event is warranted, and which could utilize multiple inner-magnetosphere and LEO observations, e.g., from THEMIS, Arase, and POES satellites, but is beyond the scope of this paper. Such an analysis could provide a comprehensive understanding of the mechanisms driving these periodic energetic electron precipitation bursts which cause periodic enhancements in X-ray flux. The authors plan to perform this investigation in a future study. 
        
        The Timepix sensors have previously demonstrated their utility for space radiation measurements, as seen for example with the SATRAM instrument onboard the Proba-V satellite \cite{Granja_2016}, and on larger CNES balloon missions, such as with the PIX instrument described by \citeA{Ruffenach_2024}. In our study, the deployment of the MiniPIX EDU on a smaller `burster' weather balloon provided high-resolution electron precipitation data at a much lower cost and required smaller logistical efforts as compared to larger balloon missions. The success reported here highlights the potential for using such compact, cost-effective Timepix-based payloads for future detailed atmospheric radiation studies perhaps at the constellation scales. The ability to capture fine details of precipitation events with minimal resources underscores the versatility and robustness of Timepix technology in diverse research environments. In terms of assessing the spatial and temporal non-uniformity of energetic electron precipitation from the Van Allen belts, a targeted launch of a small constellation of weather balloons with MiniPix-like payloads could be especially interesting.


    \section{Conclusions}
        This study highlights the successful deployment, and reports the performance of, the MiniPIX EDU sensor payload on a `burster' weather balloon. This is demonstrated by presenting detailed observations of energetic electron precipitation during the May 2024 superstorm. The main findings of the paper can be summarized as follows: 

        \begin{enumerate}
            \item The MiniPIX EDU sensor was successfully deployed on a `burster' weather balloon, capturing detailed observations of GCR radiation levels as well as energetic electron precipitation through bremsstrahlung X-rays during the May 2024 superstorm. The novel use of Timepix technology in this low-cost mission not only demonstrated the utility of the sensor for such application but also provided valuable scientific data, revealing storm-time X-ray peak structures associated with energetic electron precipitation and magnetospheric wave activity.
            \item The observed flux of GCR X-ray background was in good agreement with prior studies. Notably, the GCR electron fluxes measured by the balloon payload appeared to be approximately an order of magnitude higher during the period of the May 2024 magnetic storm than those reported in other balloon experiments \cite<e.g.,>[]{Ruffenach_2024}.
            \item The study revealed significant enhancements in X-ray flux during the storm, characterized in particular by four periodic and intense distinct peaks. These peaks in X-ray flux are attributed to bremsstrahlung emissions from energetic electron precipitation from the Van Allen radiation belt, likely modulated by Pc5 ULF waves with a $\sim$4-minute periodicity.
            \item The consistency of the observed storm-time X-ray peak structure with ground-based riometer data and satellite magnetometer measurements underscores the reliability of the MiniPIX EDU sensor's data and the effectiveness of the particle and photon identification algorithm. We show that the Timepix sensor and developed identification algorithm data can be used effectively for detecting bremsstrahlung X-rays from energetic electron precipitation from the Van Allen belts.
            \item The unexpected failure of the balloon cut-off mechanism, which extended the flight duration by an additional hour before the balloon burst, proved to be beneficial by allowing the measurement of the electron precipitation signatures during the fourth hour of the flight, underscoring that operational anomalies can sometimes yield valuable scientific insights.
        \end{enumerate}

        Overall, the novel use of Timepix sensor technology for this low-cost mission delivered valuable scientific data, revealing the storm-time measurements of modulated energetic electron precipitation in the X-ray flux data. In our view, the successful application of Timepix technology for low-cost precipitation studies on weather balloons marks a significant advancement in space weather research and has the potential for more advanced and larger campaigns perhaps at the constellation scale. Such balloon constellations could address uncertainties in the spatial and temporal characteristics of electron precipitation that are currently only poorly constrained.  
        
        
	\acknowledgments
	The balloon mission development was funded by Faculty of Engineering, and Spirit of George Ford Endowment Fund. LO is supported by a Postdoctoral Banting Fellowship. IRM is supported by a Discovery grant from Canadian NSERC. This work was partially supported by the Canadian Space Agency (CSA). CARISMA is operated by the University of Alberta, funded by the CSA as a part of Space Environment Canada. Funding for operation of the NORSTAR riometers is provided by the CSA. The authors thank M.~Mandal, N.~Klager, R.~Bererton, E.~Xiong, D.~Chan, and B.~Dundas for their help in developing the balloon payload. We also thank Stefan~Gohl for his help with the development of the particle identification algorithm. Finally, the authors thank the personnel at CFB Wainwright for their assistance in retrieving the balloon payload.  
	
	\section*{Open Research}
	OMNI solar wind data and GOES-18 magnetometer data were retrieved from the Coordinated Data Analysis Web page \cite{CDAWeb_2023}. The balloon launch raw and processed data are available from the University of Alberta repository \cite{UA_balloon_2024}.  
		
	\bibliography{biblist.bib}

\end{document}